\documentclass{iaus}
\usepackage{graphicx}

\title[Distribution of Major Semiaxes of Wide Binaries] 
{The Frequency Distribution of Major Semiaxes of Wide Binaries: Cosmogony and Dynamical Evolution}

\author[Poveda, Allen \& Hern\'andez-Alc\'antara]   
{Arcadio Poveda$^1$, Christine Allen$^1$ \& A. Hern\'{a}ndez-Alc\'{a}ntara$^1$}

\affiliation{$^1$Instituto de Astronom\'{i}a, Universidad Nacional Aut\'{o}noma de M\'{e}xico,
Ciudad Universitaria 04510 M\'{e}xico D.F. \break email: poveda@astroscu.unam.mx\\[\affilskip]}

\pubyear{2006}
\volume{240}  
\pagerange{119--126}
\date{?? and in revised form ??}

\jname{Proceedings IAU Symposium}
\editors{A.C. Editor, B.D. Editor \& C.E. Editor, eds.}

\begin{document}

\maketitle

\begin{abstract}

The frequency distribution {\it f(a)} of major semiaxes of double and multiple systems,
and their eccentricities and mass ratios, contain valuable fossil information about
the process of star formation and the dynamical history of the systems. In order to
advance in the understanding of these questions, we made an extensive analysis of the
frequency distribution {\it f(a)} for wide binaries ($a>25$ AU) in various published catalogues,
as well as in our own (Poveda \etal\ 1994; Allen \etal\ 2000;
Poveda \& Hern\'{a}ndez-Alc\'{a}ntara 2003). Based upon all these studies 
we have established that the frequency distribution {\it f(a)} is function of the age of the 
system and follows
Oepik's distribution {\it f(a)} $ \sim $  1/{\it a} in the
range of 100 AU $<$ {\it a} $<$ {\it a}$_{c}$({\it t}), where {\it 
a}$_{c}$({\it t})
critical  major semiaxis beyond which binaries have been dissociated by 
encounters
with massive objects. We argue that the physics behind the distribution 
{\it f(a)} $ \sim $
1/{\it a} is a process of energy relaxation, analogous to those present in 
stellar clusters
(secular relaxation) or in the early stages of spherical galaxies (violent 
relaxation).
The existence of runaway stars indicates that both types of relaxation are 
important
in the process of binary and multiple star dynamical evolution.


\end{abstract}

\firstsection 
\section{Introduction}

The distribution of major semiaxes (separations) of double and multiple 
stars is a
fossil record of the conditions at star formation, as well as of the 
processes of
dynamical evolution, including the dissociation of wide binaries produced=
 by
encounters with massive objects: molecular clouds, spiral arms, etc.
Our long-standing interest in these topics has led us to investigate the 
frequency
distribution of major semiaxes (separations) of wide binaries as a 
function of age.
In the past, two main distributions of major semiaxes have been proposed:

$ \quad $(1) a power-law frequency distribution {\it f(a)}   $ \sim 
$  {\it a}$^{-\alpha}$

$ \quad $(2) a Gaussian distribution in log {\it P} or log {\it a}.

When   $ \alpha $ = 1 in the power-law distribution, we have the well 
known Oepik (1924)
distribution (OD). On the other hand, Kuiper (1935, 1942) proposed the 
Gaussian
distribution, which was further elaborated by Heintz (1969). More recently,
Duquennoy \& Mayor (1991, DM) again proposed a Gaussian distribution in 
log {\it P}, valid
throughout the interval 1 $<$ log {\it P} (days) $<$ 10.

         Our interest in the subject led us to construct a catalogue of=
 wide
binaries in the solar vicinity, based on the catalogue of nearby stars of 
Gliese and Jahreiss
(1991, GJ); see Poveda \etal\ (1994) for details. We have also constructed=
 a
list of common proper motion binaries in the Orion Nebula Cluster (age 
10$^{6}$ years), extracted
from the Jones \& Walker (1988) catalogue of proper motions. (Poveda \& 
Hern\'{a}ndez-Alc\'{a}ntara
2003).
In our search for evolutionary effects in the observed distributions we 
have also looked at the
oldest stars in the Galaxy. For this purpose we constructed a catalogue 
(Allen  \etal\ 2000) of
common proper motion companions to the lists of high velocity metal-poor 
stars of Schuster \etal\
(1988; 1989a; 1989b), with ages of about $10^{10}$ years.

         In all our catalogues, as well as in the Luyten Double Star 
 Catalogue
(1940-1987, LDS) and in Chanam\'{e} \& Gould's Catalogue (2004, CG), we 
confirm our previous
findings
(Poveda \etal\ 1997; Poveda \& Allen 2004), that the separations follow 
Oepik's distribution (OD)
in an interval that is bounded at the lower end ({\it a} $ \sim $ 100 AU) 
by the process of
close binary and
  protoplanetary disk formation, and at large separations ($a >$ 2500 AU) 
 by the dissociation effects
  produced by encounters with massive objects.

         It can be shown that Oepik's distribution in the plane ({\it N}, 
 log {\it P})
is a horizontal straight line which is quite consistent (within its error 
bars)
with DM distribution in the interval 2.44 $<$ log {\it P} (years) $<$ 
5.44, which corresponds
to 53 $<$ {\it a} (AU) $<$ 5500. (Poveda  \etal\ 2004). Since a great 
number of binaries
from many different and largely independent sources
confirm the validity of Oepik's distribution, and since there is no 
stellar formation or single
physical
process able to produce a Gaussian distribution valid in the
interval 1 $<$ log {\it P} (days) $<$ 10,
we propose
to abandon the Gaussian representation for {\it a} $>$ 100 AU. On the 
contrary, Oepik's
distribution, which is
equivalent to a surface density of secondaries $\rho $({\it a})  $ \sim $ {\it a}$^{-2}$,
has a physical interpretation. In fact,
this distribution is similar to the run of surface brightness in globular 
clusters (King 1962) or
to that of elliptical galaxies (Hubble's 1930, or de Vaucouleurs'=92 1953 
distributions).
In both cases the physics
behind such distributions is well known: it is the result of energy 
relaxation. The similarity
of OD to the surface brightness in clusters and elliptical galaxies 
indicates that binaries are
not born alone;
at birth, they must be subject to a process of energy relaxation which
cannot be produced by two-body encounters, i.e., stars must be formed in 
groups of
multiplicity $n \ge  3$.

This paper is organized as follows.  In Section 2 we study the 
distribution {\it f(a)}
for a volume-complete
sample extracted from our catalogue (Poveda \etal\ 1994). Section 3=
 examines
the distribution {\it f(a)} in our catalogue of very young ({\it T} $<$ 
10$^{6}$ years),
common proper motion binaries in the Orion Nebula Cluster (Poveda \&
Hern\'{a}ndez-Alc\'{a}ntara 2003).
In Section 4 we examine another sample of wide
binaries, namely Chanam\'{e} \& Gould\'s (2004) common proper motion 
binaries from the
revised NLTT (Salim \& Gould 2003; Gould \& Salim 2003). Again, the
binaries in this catalogue follow OD. Section 5 examines the physics 
behind Oepik's
distribution,
and Section 6 presents our conclusions.


\begin{figure}[!ht]
\vspace{0.7cm}
\centerline{\includegraphics[height=2.2in,width=3.3in]{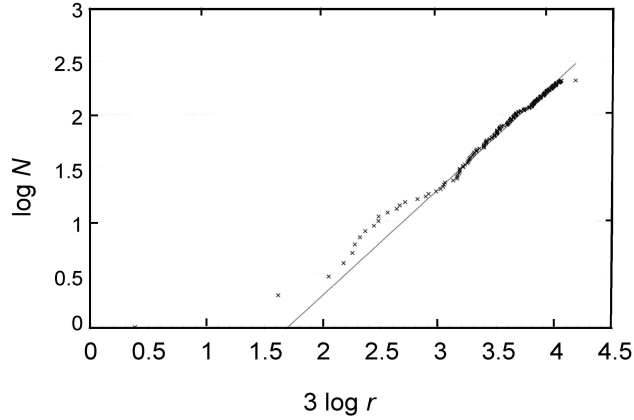}}
\caption{Completeness of the systems in the catalogue
of nearby wide binary and multiple systems (Poveda  \etal\ 1994).
The straight line corresponds to $N(r) \sim\ r^{3}$, i.e., a 
volume-complete sample.}
\label{fig1}
\vspace{0.25cm}
\end{figure}


\begin{figure}[!ht]
\centerline{\includegraphics[height=2.2in,width=4.8in]{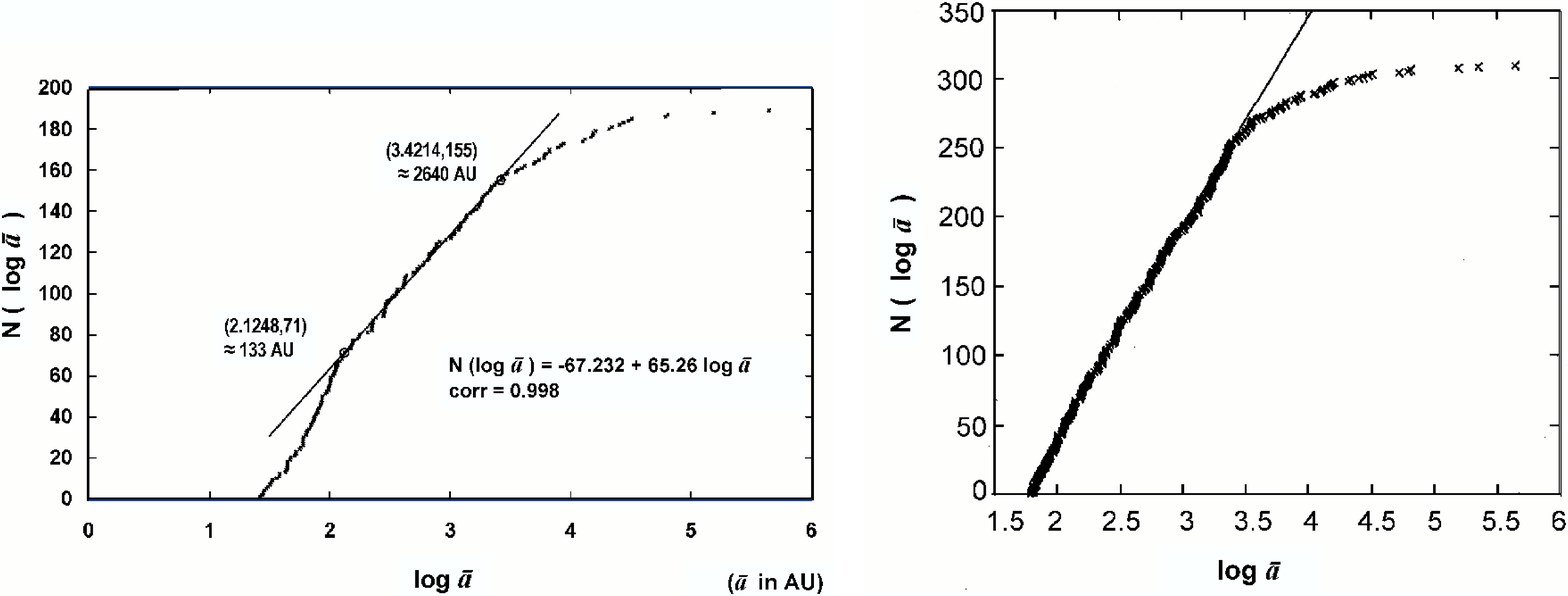}}
\caption{(a)Left. Cumulative distribution of $log a$ for the 189 binaries 
of the complete sample
in Poveda \etal\ (1994), ie, for systems with $M_v < 9$ and  luminosity 
class IV-V.
The straight ine is a fit for Oepik's relation.  The KS test for this fit 
gives a value of
$Q= 0.99$ in the interval (133AU, 2640AU). (b)Right. Comparison between 
Oepik's distribution and
the cumulative numbers of log $a$ for the binaries in the Poveda \etal\ 
(2004) catalogue.  The
straight line is a fit to Oepik's distribution for systems 
satisfying  $60  < a$ (AU) $< 2965$.
The KS test gives a value of $Q = 0.96$ for the interval $60  < a$ (AU)=
 $< 
2965$.}
\label{fig2}
\vspace{0.15cm}
\end{figure}

\section{A Catalogue of Nearby Wide Binaries and a Volume-Complete Sample 
of these Objects.}\

With the purpose of detecting the effects of dissociation of weakly bound 
binaries with
the passage of time, we constructed a catalogue of wide binaries (305 
double systems,
26 triples and 3 quadruples with $a >$ 25 AU), (Poveda  \etal\ 1994).

    A sub-sample of this catalogue i.e., all those systems with primaries 
 of luminosity
class V or IV and brighter than absolute magnitude $M_{\it v} = $ 9 is 
very important, as it is
volume-complete. To show this, in Figure 1 we plot {\it N}(log {\it r}) 
vs. 3 log {\it r}, where
{\it N}(log {\it r})
is the number of systems ($M_{\it v} < $9) out to a distance  {\it r}. As 
can be
seen, with the exception
of a few very close systems, the great majority follow the relation {\it 
N} $ \sim $ {\it r}$^{3}$
right to
the limit of the catalogue ({\it r} = 22.5 pc), as expected for a 
volume-complete sample.

Having a volume-complete catalogue, we proceed to investigate the 
frequency distribution
of major semiaxes. It can be shown that an equivalent representation of OD 
is the cumulative
distribution {\it N}($<$ log {\it a}) $\sim$ log {\it a}. In the
plane {\it N}($<$ log {\it a}) - log $a$,
OD is a straight line. In general,
we will favor the analysis of the cumulative distribution {\it f(a)} in 
order to
reduce the noise introduced by small number sampling fluctuations. Since 
most of the wide
binaries (in our catalogues) do not have reliable orbits because of their 
long periods,
we have to use angular separations {\it s} and distances to give projected 
separations in
astronomical units. However, a statistical relation between the average 
value of {\it a} for
a given projected separation {\it s}, namely $\bar{a}$ = 1.41{\it s} 
(Couteau 1960), can be used to
estimate for each binary a value of $\bar{a}$.

Figure 2a shows for our volume-complete sample the run of the cumulative 
distribution
$N (log \bar{a}) vs. log \bar{a}$ . An inspection of Figure 2a shows that 
this sample closely
follows Oepik's distribution in the interval
2.12 $<$ log $\bar{a}$ $<$ 3.42, i.e., 133 $<$ $\bar{a}$ (AU) $<$ 2,640.
To evaluate
quantitatively how reliably Oepik=B4s distribution represents the data, we 
shall use the
Kolmogorov-Smirnov (KS) test, which was developed precisely for cumulative 
distributions
and where no arbitrary binning of the data is required, (as is the case in 
the popular
Chi-square test). In Figure 2a each binary is plotted, thus {\it N}(log 
$\bar{a}$)  increases
one by one.
By least squares we fitted to the data points a number of straight lines, 
each one defined
in the interval (log $\bar{a}_{i}$, log $\bar{a}_{j}$), until we found the 
best straight
line that minimized the
residuals and maximized the interval (log $\bar{a}_{i}$, log 
$\bar{a}_{j}$), i.e., the one giving
the largest interval
(log $\bar{a}_{i}$, log $\bar{a}_{j}$) in which Oepik's distribution 
reliably represents the
cumulative distribution of separations.

In Figure 2a we give the equation of the straight line that best 
represents the data,
in the interval log $\bar{a}_{i}$ = 2.1248, log $\bar{a}_{j}$ = 3.4214 
($\bar{a}_{i}$ $\approx$ 133
AU, $\bar{a}_{j}$ $\approx $ 2,640 AU). Having found
Oepik's distribution for the wide binaries in the interval 133 $<$ 
$\bar{a}$(AU) $<$ 2640 we now
proceed to test, via KS, what is the level of significance of the 
theoretical distribution OD.
The closer to 1 the estimator $Q$ is, the better the theoretical 
representation (Press \etal\ 1990).
For the
present case we find {\it Q} = 0.99, i.e.,
we can accept Oepik's distribution at a very high level of confidence in 
the interval (133 AU,
2640 AU).
For the interval (133 AU, 3100 AU) we find $Q = 0.96$, also representing=
 a 
high level
of confidence.

In Figure 2b, we plot in the same plane as in Figure 2a, all the binaries 
(305)
from our 1994 catalogue. Even though this sample is not volume-complete, 
one can
argue that this does not seem to introduce an important bias in the 
distribution
of separations. In fact, as can be seen from Figure 2b, the cumulative 
distribution
{\it N}($<$ log $\bar{a}$) again defines a straight line. Repeating the 
statistical
analysis for this catalogue of 305 binaries we find that the KS test gives 
a value
{\it Q} = 0.96 for $\bar{a}$ in the interval between 60 AU and 2965 AU.

We now seek an explanation for the limits of validity of the OD as shown 
in Figures 2a and 2b,
particularly for the volume-complete sample. The following hypothesis is 
proposed:
(1) at the short end of the distribution $\bar{a}$ $\sim $ 100 AU), any 
primeval OD will be quickly
modified by the presence of close binaries and protoplanetary disks; (2) 
at the wide
end ($\bar{a}$ $>$ 3,000 AU),
encounters with massive objects (molecular clouds, spiral arms, black 
holes, MACHOS, etc.) will
gradually dissociate the widest binaries. In fact, in a theoretical paper, 
Weinberg \etal\ (1987)
estimated that a binary with a major semiaxis smaller than 2000 AU has a 
probability of one to
survive (against dissociation by giant molecular clouds) during 10 billion 
years, yet a binary
with $\bar{a}$ $\approx$ 12,600 AU after two billion years has a 
probability of survival of
only 0.5.
The consistency between the results of Weinberg \etal\ and the 
distributions shown in Figs. 2a and
2b gives support to the hypothesis that Oepik's distribution is primeval, 
valid up
to $\bar{a}$ $\approx$ 45,000 AU (see next section),
but with the passage of time it gets truncated at the large separations.
  To further examine this hypothesis, we analyze the distribution of major 
 semiaxes of the binaries
in a very young group ($T< 10^{6}$ years).


\begin{figure}[!ht]
\vspace{0.3cm}
\centerline{\includegraphics[height=2.2in,width=3.3in]{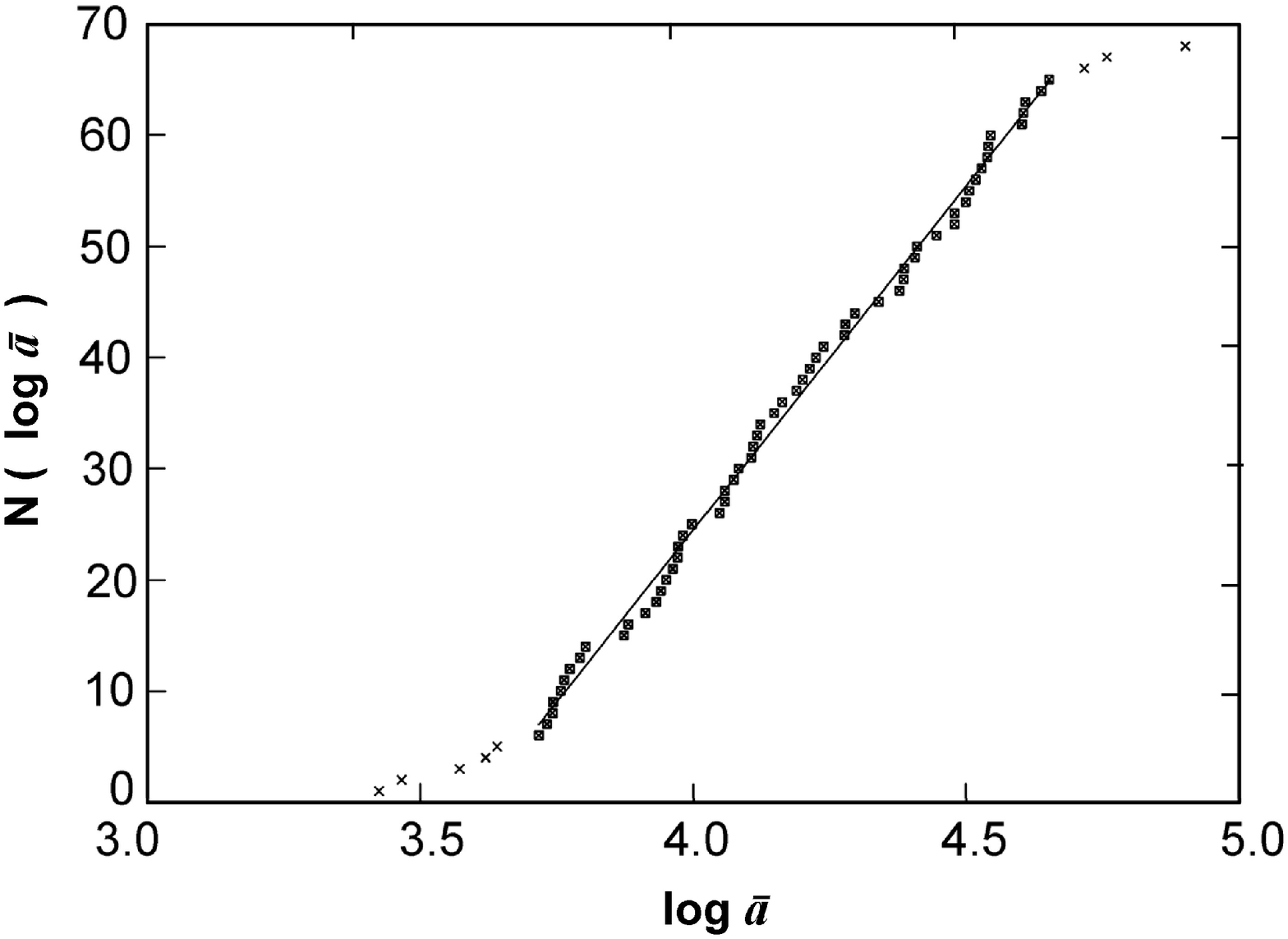}}
\caption{Cumulative distribution of the logarithms of the expected value 
of a for the common
proper motion binaries of the Orion Nebula cluster (Poveda \& 
Hern\'andez-Alc\'antara 2003).
The straight line represents Oepik's distribution.  For the interval 
fitted with a straight line,
$5180 < \bar{a} $(AU) $< 44800$ (60 binaries) the KS test gives a value of 
$Q = 0.99$.}
\label{fig3}
\end{figure}

\section{Wide Binaries in a Very Young Group}

Taking advantage of the Jones \& Walker (1988) catalogue of proper motions 
of the stars in the Orion
Nebula Cluster, we have identified 68 candidate common proper motion=
 binaries
(Poveda \& Hern\'{a}ndez-Alc\'{a}ntara, 2003). Jones \& Walker's 
determinations of proper
motions and infrared magnitudes for 1053 stars of the Orion Nebula Cluster 
are
appropriate for our work, because of the proper motions accuracy
($\sigma_{\mu}~<$~0.1~arcsec/century)
and also because the faint limiting magnitude ($I <$ 13). Moreover, the 
one- million year
age of the cluster allows us to look into the =93almost=94 primeval 
distribution of major
semiaxes. The cumulative distribution of major semiaxes for this sample of 
Orion binaries
follows very neatly Oepik's distribution, (see Figure 3). The KS analysis 
of the data plotted
in Figure 3 indicates that OD fits the data up to major semiaxes of 45,000 
AU with a $Q$ = 0.99.


\begin{figure}[!h]
\vspace{0.2cm}
\centerline{\includegraphics[height=2.2in,width=3.3in]{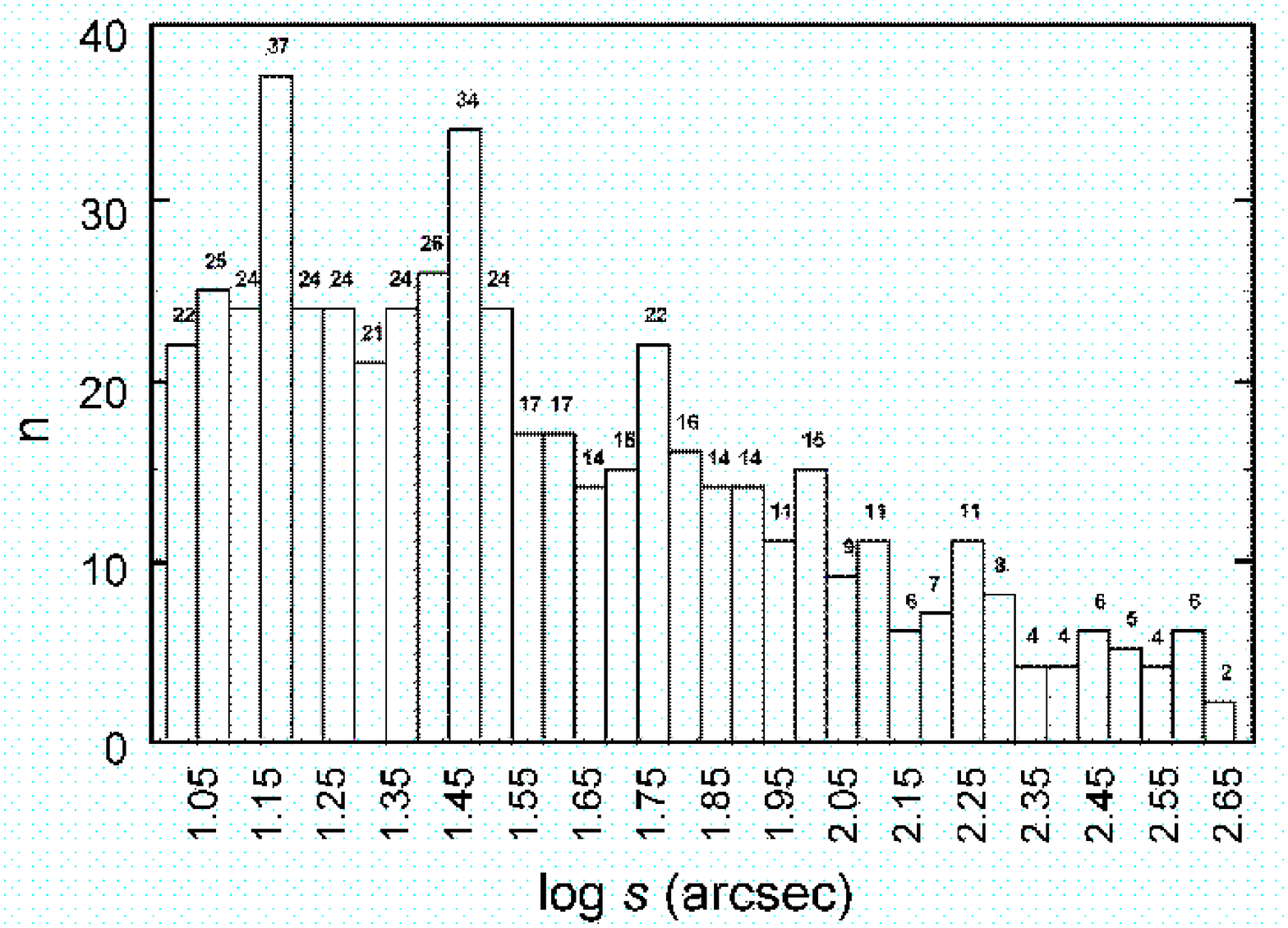}}
\caption{
Frequency distribution of the separations s for the disk binaries of 
Chanam\'e \& Gould (2004) for
binaries with $s > 10$ arcsec (523 binaries). Note that in the interval $1 
< $ log $s < 1.55$
(10 arcsec $< s < 36$ arcsec)
which includes 285 binaries, this distribution is consistent with Oepik's, 
which in this plane
would be a horizontal straight line. For larger separations, the 
distribution becomes
depopulated similarly to what we found to occur for other samples of 
binaries (see text).
We interpret this departure as due to the dissociation of wide binaries by 
encounters with
massive perturbers.}
\label{fig4}
\vspace{0.3cm}
\end{figure}

\section{Another Sample of Wide Binaries.}

Chanam\'{e} and Gould (2004, CG) have assembled a catalogue of wide=
 binaries
based on the revised Luyten NLTT Catalogue by Gould \& Salim (2003) and
Salim \& Gould (2003). CG identified 999 common proper motion pairs. Good 
photometry
allowed them to construct a reduced  proper motion diagram for their 
binaries.
The position of the binary components in this diagram helps to separate=
 disk
main-sequence pairs (801) from halo subdwarfs (116). The large number of 
binaries
  in the CG Catalogue offers an independent sample to test the validity of 
 OD for
main-sequence and halo binaries, respectively.

According to CG, their catalogue is incomplete for separations {\it s} $<$ 
10 arcsec.
Since we are trying to establish the frequency distribution of separations 
of binaries in
  CG we extracted from their list a sample of disk binaries with=
 separations
{\it s} $>$ 10 arcsec, i.e.,
we rejected 276 binaries closer than 10 arcsec out of 800 disk 
binaries.  For the remaining 524
pairs we plot in Figure 4 the frequency distribution of separations 
$f(s)ds$. The large number
of binaries allows to display this frequency distribution with small 
sampling fluctuations.
In this figure we see clearly that $f(s)ds$ is essentially constant in the 
interval 1 $\leq $
log {\it s} $\leq $ 1.55. For larger separations, i.e., {\it s} $ >$ 35.5 
arcsec,  $f(s)$
  is a decreasing function of ${\it s}$.  CG noticed this behavior, which 
 they found
  statistically significant. The constancy of
$f(s)$ in the interval 10 $<$ {\it s} (arcsec) $<$ 35.5 is just what we 
expect from OD. Since
CG estimate the
mean distance of their disk binaries to be 60 pc, we can transform into 
astronomical units
the separations listed by CG. The angular interval where OD holds 
transforms into:
600 $\leq s$ (AU) $\leq$ 2,129, which is equivalent to 840 $\leq$ 
$\bar{a}$ (AU) $\leq $ 2,981.


\begin{figure}[!ht]
\vspace{0.4cm}
\centerline{\includegraphics[height=2.2in,width=2in]{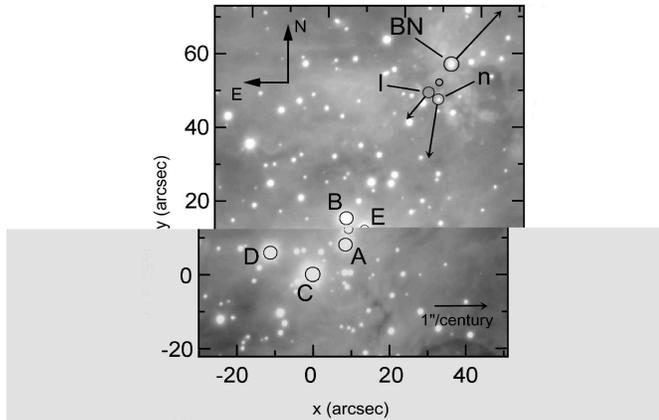}}
\caption{An example of a recent and nearby case of violent relaxation
(Rodr\'iguez \etal\ 2005;  G\'omez \etal\ 2006) Next to the Orion
Trapezium, in the Becklin-Neugebauer, Kleinman-Low region we show 3 
sources: BN, $I$, $n$,
which are moving away from a center with transverse velocities of the 
order of 26, 8 and
28 km s$^{-1}$, respectively. These velocities are much larger than those 
expected for secular
relaxation. Indeed, they correspond to violent relaxation.}
\label{fig5}
\end{figure}

\section{The Physics Behind Oepik's Distribution}

The analysis of the various samples of wide binaries as shown in Figs. 
2-5, show convincingly
  that Oepik's distribution holds for {\it a} $>$ 100 AU.  The distribution
{\it f(a)} $ \sim a^{-1}$ is equivalent to
a surface density of secondaries  $\rho$({\it a}) $ \sim a^{-2}$. This 
surface density
distribution is reminiscent
of the surface density of stars in globular clusters and elliptical=
 galaxies
(King 1962; Hubble 1930; de Vaucouleurs 1953). The physics involved in 
these distributions
is well understood. In the case of globular clusters we have what we may 
call secular
relaxation, i.e., a process in which two body encounters gradually modify 
the energy of
a given star over a time scale much greater than the crossing time; in the 
case of
elliptical galaxies, what we have is a case of initially violent=
 relaxation,
in which the energy of the stars changes rapidly because the potential of=
 the
stellar system (galaxy) changes rapidly due to an initial stellar collapse
(van Albada 1982) or because of galaxy mergers. In the first case (secular 
relaxation)
there is equipartition of energy and therefore a tendency for the lighter 
stars to
diffuse to the outer parts of the globular cluster (this has been 
observationally confirmed);
on the contrary, in the case of violent relaxation there is no 
equipartition of energy and hence
no segregation of stellar masses, which indeed is the case in elliptical 
galaxies (no color
gradients are observed).

         The process of energy relaxation involves the interaction of 
 several close stars
({\it n} $\geq$ 3), very early in the stellar history. In fact, the wide 
binaries in the Orion
Nebula Cluster follow OD even though they are still in the pre-main 
sequence phase
(T $<$ 10$^{6}$ years).
We conclude that the process of energy relaxation is not related to 
interactions with stars
in the cluster environment, but rather it is the result of very early 
interactions in  multiple
systems.
This suggests that stars are formed in multiple systems that quickly relax 
and assume OD, with
the possible
ejection of one or more single stars. The common proper motion binaries in 
Orion show that there
is not enough time for the reverse process to take place, i.e. the 
formation of double and multiple
  stars by capture from stars in the cluster.

         In the process of star formation, the transition from gas to 
 stellar dynamics may lead to a
virialized multiple system where 2{\it T} $ + \Omega \approx 0$, or to a 
non virialized one,
depending on the initial conditions
prevailing in the transition phase. If initial condensations (proto-stars) 
are not virialized,
2{\it T} + $\Omega \ll  0$, then the proto-stars will collapse towards the 
center of mass of the
multiple systems.
Here we meet the conditions of violent relaxation, i.e., the collapse of 
the proto-stars will
produce
a rapid change in the gravitational potential experienced by the stars. In 
this conditions some
stars
will be accelerated to velocities larger than those associated with a 
virialized multiple system.
An {\it n}-body simulation of this scenario was realized by the present 
authors many years ago
(Poveda \etal\ 1967) with the purpose of finding an alternative 
explanation for the formation
of runaway stars. In those simulations we found not only runaway stars but 
also that the
binaries formed followed Oepik's distribution (Allen 1968; Poveda  \etal\ 
2004).
Figure 5 shows a recent and nearby case of violent relaxation
(Rodr\'{i}guez  \etal\ 2005; G\'{o}mez \etal\ 2006).
In the vicinity of the Orion Trapezium, in the Orion molecular cloud, the 
proper
motions of the heavily obscured B star (Becklin-Neugebauer Object), as 
well as of the infrared
objects $I$ and {\it n} imply transverse velocities much larger than those 
expected to be produced
in a virialized multiple system.

\section{Conclusions}\label{sec:concl}

(1) The study of a large number of wide binaries, mostly from independent 
sources
confirms that the frequency distribution of major semiaxes is {\it f(a)} $ 
\sim a^{-1}$, i.e.
precisely
Oepik's distribution. This distribution is truncated at the short end ($a 
\approx $ 100 AU) by the
presence of close binaries and protoplanetary disks; at the large 
separations the
distribution is depopulated by the process of dissociation produced by 
encounters with
massive objects. Figure 2 clearly exhibits these effects for the 
volume-complete sample
in the solar vicinity.

(2)  Oepik's distribution in the plane {\it N} (log {\it P}) is a 
constant, which is entirely
consistent (within the error bars) with Duquennoy \& Major's Gaussian 
distribution
in the interval 42$<$ $\bar{a}$ (AU) $<$ 4213.

(3)  Since there is no single astrophysical process that would generate a 
Gaussian
distribution in log {\it P} (or in log {\it a}) over such a wide interval
(1 day $<$ {\it P} $< 10^{10}$ days)
and since, on the other hand, Oepik's distribution can be explained by the 
process of energy
relaxation in  few-body interactions, we propose to abandon the Gaussian 
representation
for log {\it a}, in favor of Oepik's distribution (for {\it a} $>$ 100 AU.)

(4)  Oepik's distribution suggests that the process of star formation 
produces multiple
stars which evolve towards binaries after ejecting one or more single=
 stars.

Acknowledgement.  Our thanks are due to Paola Ronquillo for her help in 
the preparation of the
typescript.

\end{document}